# DATA QUALITY MEASUREMENT ON CATEGORICAL DATA USING GENETIC ALGORITHM

J.Malar Vizhi[1] and Dr. T.Bhuvaneswari[2]

[1]Research scholar,Bharathiar University,Coimbatore,Tamil Nadu,India.
jmalarvizhi@lycos.com
[2]Assistant Professor in the Department of Computer Science, Government College for women, Burgoor,Tamil Nadu,India.
t_bhuvaneswari@yahoo.com

***ABSTRACT***

*Data quality on categorical attribute is a difficult problem that has not received as much attention as numerical counterpart. Our basic idea is to employ association rule for the purpose of data quality measurement. Strong rule generation is an important area of data mining. Association rule mining problems can be considered as a multi objective problem rather than as a single objective one. The main area of concentration was the rules generated by association rule mining using genetic algorithm. The advantage of using genetic algorithm is to discover high level prediction rules is that they perform a global search and cope better with attribute interaction than the greedy rule induction algorithm often used in data mining. Genetic algorithm based approach utilizes the linkage between association rule and feature selection. In this paper, we put forward a Multi objective genetic algorithm approach for data quality on categorical attributes. The result shows that our approach is outperformed by the objectives like accuracy, completeness, comprehensibility and interestingness.*

## 1. INTRODUCTION

Data Mining is the most instrumental tool in discovering knowledge from transactions [1, 2].The most important application of data mining is discovering association rules. This is one of the most important methods for pattern recognition in unsupervised systems. These methods find all possible rules in the database. Some measures like support and confidence are used to indicate high quality rules.

Data mining techniques can be employed in order to improve Data quality. Poor data quality is always a problem in practical application. Data Quality Mining can be defined as the deliberate applications of Data Mining techniques for the purpose of Data quality, measurement and improvement [3].

Data Quality Mining is one of the most important tasks in data mining community and is an active research area because the data being generated and stored in databases of organizations are already enormous and continues to grow very fast. This large amount of stored data normally contains valuable hidden knowledge, which if harnessed could be used to improve the decision making process of an organization. However the volume of the archival data often exceeds several gigabytes and sometimes even terabytes. Such an enormous volume of data is beyond the manual analysis capability of human beings. Thus, there is a clear need for developing automatic methods for extracting knowledge from data that not only has a high accuracy but also comprehensibly and interestingness by user.

We describe a first approach to employ Association Rules for the purpose of data quality mining. Data mining algorithms like Association Rule Mining perform an exhaustive search to find all rules satisfying some constraints [4].

Genetic Algorithm is a new approach for Association rule mining. Finding the frequent rules is the most resource consuming phase in association rule mining and always does some extra comparison against the whole database [5].Although Genetic Algorithm is good at searching for undetermined solution, it is still rare to see that Genetic Algorithm, is used to mine Association Rules. By introducing Genetic Algorithm on the rules generated by Association Rule, it is based on four Objectives Accuracy, Completeness, Comprehensibility and Interestingness. Data Quality Mining using Genetic Algorithm is applied for categorical attributes. It is a widely used technique in which data points are partitioned into groups in such a way that points in the same group.

Nowadays improvement in techniques allows shop to collect several data about customer's market baskets; a basket indicates items purchased by a customer at a specific time. Categorical attributes usually have small domains. Typical categorical domain considered for grouping consists of less than a hundred or, rarely a thousand attribute Values.

An important character of categorical attributes is that they typically have a small number of attribute values. Quantitative attributes can be discretizing into categorical attributes.

For example: if age is discretized into steps of 2 years we would probably find rules
Age(X, 18...19) and lives (X, Lausanne) → profession(X, student)
Age(X, 20...21) and lives (X, Lausanne) → profession(X, student)
Could be also expressed as a rule
Age(X, 18...21) and lives (X, Lausanne) → profession(X, student)
This is more compact but requires a different discretization.

## 2. ASSOCIATION RULES

The process of discovering an interesting and an unexpected rule from large data sets is known as Association rule mining. Association rule mining is one of the most challenging areas of data mining which was introduced in Agrawal to discover the associations or co-occurrence among the different attributes of the dataset. Several algorithms like Apriori(Agrawal), SETM(Houtsma and Swami), AprioriTID(Agrawal and Srikant), DIC(Brin), Partition algorithm(Savasere), Pincer search(Lin and Kedem), FP-tree(Han) etc have been developed to meet the requirements of this problem. These algorithms work basically in two phases: the frequent itemset generation and the rule generation. Since the first phase is the most time consuming, all the above mentioned algorithms maintain focus on the first phase. A set of attributes is termed as frequent set if the occurrence of the set within the dataset is more than a user specified threshold called minimum support. After discovering the frequent itemsets, in the second phase rules are generated with the help of another user parameter called minimum confidence.

The typical approach is to make strong simplifying assumption about the form of the rules and limit the measure of rule quality to simple properties such as support and confidence [6]. Support and confidence limit the level of interestingness of the generated rules. Completeness and accuracy are metrics that can be used together to find interesting association rules .The major aim of Association rule mining is to find the sets of all subsets of items or attributes that frequently occur in many databases records or transaction. Association Rule Mining algorithm discover high level prediction rules in the form

IF THE CONDITION of the values of the predicating attributes are TRUE
THEN
Predict values for some goal attributes.

An Association Rule is a conditional implication among itemsets $X \rightarrow Y$ where $X \rightarrow Y$ are itemsets and $X \cap Y = \Phi$. An itemset is said to be frequent or large if it support is more than a user specified minimum support value. The confidence of an Association Rule is given as Support (XUY)/Support(X) is the conditional probability that a transaction contains Y given that is also contains X.

## 3.CATEGORICAL ATTRIBUTES

Categorical attributes, whose domain is not numeric is a difficult yet important task to many fields from statistics to psychology deal with categorical data. Increasingly, the data mining community is inundated with a large collection of categorical data. Grouping mixed data sets into meaningful groups is a challenging task in which a good distance measure, which can adequately capture data similarities, has to be used in conjunction with an efficient grouping algorithm. In order to handle mixed numeric and categorical data, some of the strategies that have been employed are as follows.

(1) The numeric distance measures can be applied for computing similarity between object pairs after conversion of categorical and nominal attribute values to numeric integer values.

(2) Another approach has been to propose discretize numeric attributes and apply categorical clustering algorithm. But the discretization process leads to loss of information.[7]

An important implication of the compactness of categorical domain is that the inter attribute $\sum_{ij}$ for any pair of attributes $A_i$ and $A_j$ fits into main memory because the number of all possible attributes value pairs from $A_i$ and $A_j$ equals $|D_i|*|D_j|$ where A and D be the attributes and domain.

Let $A_i$ be an attributes with an usually large domain Di without loss of generality, let $D_i$ be the set $\{1\ldots|D_i|\}$, Let $M<|D_i|$ be the maximum number of attribute values per attribute so that categorical attribute fits into main memory. Let $C= [|D_i|/M]$.We construct $D_i$ of size M from $D_i$ by mapping for a given $x\in (0,\ldots, M-1)$, the set of attribute value $\{x.c+1,\ldots,x.c+c\}$ to the value x+1.[8] Formally $D_i=\{ \quad f(1),...,f(|D_i|)\}$ where $f(i)=[i/k]+1$ .

Most of the earlier works on numeric attributes which have a natural ordering on their attributes values but categorical attributes do not have a natural ordering. Each tuple in the dataset describes a sample of gilled mushroom using 22 categorical attributes. [9]

For example:
Cap-shape: =bell, conical, convex, flat, knobbed, sunken
gill-color:=black,brown,buff,chocolate,gray,green,orange,pink,purple,red,white,yellow

Expectation maximization can also accommodate categorical variables. At first, the method will randomly assign different probabilities to each class or category for each group. In successive iteration these probabilities are refined to maximize the likelihood of the data given the specified number of groups.

Handling Categorical Attributes

- Potential Issues
  - What if attribute has many possible values.

- Many of the attribute values may have very low support
  - Potential solution: Aggregate the low-support attribute values
  - What if distribution of attribute values is highly skewed.
  - Potential solution: drop the highly frequent items

A typical approach for handling categorical attributes is to transform them into asymmetric binary variables so that existing association rule mining algorithms can be applied to the data set. Categorical attributes are transformed into asymmetric binary variables by introducing as many new “items” as the number of distinct attribute-value pairs.

Table 1. Example for mining association rules.

| **Session Id** | **Country** | **Session length** | **Number of pages viewed** | **Gender** | **Browser type** | **Buy** |
|---|---|---|---|---|---|---|
| 1 | USA | 982 | 8 | Male | Internet Explorer | No |
| 2 | China | 811 | 10 | Female | Netscape | No |
| 3 | USA | 2125 | 45 | Female | Mozilla | Yes |
| 4 | Germany | 123 | 9 | Male | Internet Explorer | Yes |

For example, the categorical attribute Browser Type can be replaced by asymmetric binary variables such as Browser=Internet Explorer, Browser=Netscape, and Browser=Mozilla, while the attribute Country of Origin can be replaced by the names of all countries present in the data set. Even a symmetric binary attribute such as Gender can be handled in this way by replacing it with the asymmetric binary variables Male and Female. Table 2 shows the results of binarizing the categorical and symmetric binary attributes of the Web data. De-spite the addition of new items, the width of each transaction remains unchanged as each Web user has only one country of origin, browser type, and gender. However, one problem with this approach is that some categorical attributes such as Country of Origin can have a large number of distinct values
Creating a new item for each country name may increase the dimensionality of the Web data tremendously because there are more than 190 independent countries around the world. If the support threshold is low enough such that most of the newly created items are frequent, the problem becomes compute intensive since a large number of candidate itemsets must be generated from these frequent items. (Nevertheless, the maximum size of frequent itemsets found does not change even with the addition of new asymmetric binary variables because it depends only on the maximum width of the transactions)[10].

Table 2. Example of binarizing the categorical attributes**.**

| Session Id | USA | China | Germany | Male | Female | Internet Explore r | Nets cape | Mozilla |
|---|---|---|---|---|---|---|---|---|
| 1 | 1 | 0 | 0 | 1 | 0 | 1 | 0 | 0 |
| 2 | 0 | 1 | 0 | 0 | 1 | 0 | 1 | 0 |
| 3 | 1 | 0 | 0 | 0 | 1 | 0 | 0 | 1 |
| 4 | 0 | 0 | 1 | 1 | 0 | 1 | 0 | 0 |

# 4. GENETIC ALGORITHM

Genetic Algorithm (GA) was developed by Holland in 1970. This incorporates Darwinian evolutionary theory with sexual reproduction. Genetic Algorithm is stochastic search algorithm modeled on the process of natural selection, which underlines biological evolution. Genetic Algorithm has been successfully applied in many search, optimization, and machine learning problems. Genetic Algorithm process works in an iteration manner by generating new populations of strings from old ones.

Every string is the encoded binary, real etc., version of a candidate solution. Standard Genetic Algorithm apply genetic operators such selection, crossover and mutation on an initially random population in order to compute a whole generation of new strings.

• Selection deals with the probabilistic survival of the fittest, in that fit chromosome are chosen to survive, where fitness is a comparable measure of how well a chromosome solves the problem at hand.

• Crossover takes individual chromosomes from Parents, combines them to form new ones.

• Mutation alters the new solutions so as to add stochastic in the search for better solutions.

In general the main motivation for using Genetic Algorithms in the discovery of high-level prediction rules is that they perform a global search and cope better with attribute interaction than the greedy rule induction algorithms often used in data mining. This section discusses the aspects of Genetic Algorithms for rule discovery.

## 4.1 Genetic Operators for Rule Discovery

There have been several proposals of genetic operators designed particularly for rule discovery. Although these genetic operators have been used mainly in the classification task, in general they can be also used in other tasks that involve rule discovery, such as dependence modeling. We review some of these operators in the following subsections.

### 4.1.1 Selection

In the Michigan approach, each individual represents a single rule. Since the goal of the algorithm is to discover a set of (rather than just one) rules, it is necessary to avoid the convergence of the population to a single individual (rule). The Algorithm does that by using a selection procedure called universal suffrage. In essence, individuals to be mated are “elected” by training examples. Only rules covering the same examples compete with each other.

#### 4.1.1.1 Michigan versus Pittsburgh Approach

Genetic algorithms (Genetic Algorithms) for rule discovery can be divided into two broad approaches, based on how rules are encoded in the population of individuals ("chromosomes"). In the Michigan approach each individual encodes a *single* prediction rule, whereas in the Pittsburgh approach each individual encodes *a set of* prediction rules[11].

We use the term "Michigan approach" in a broader sense, to denote any approach where each Genetic Algorithm individual encodes a single prediction rule, the choice between these two approaches strongly depends on which kind of rule we want to discover. This is related to which kind of data mining task we are addressing. Suppose the task is classification. Then we usually evaluate the quality of the rule set as a whole, rather than the quality of a single rule. In other words, the interaction among the rules is important. In this case, the Pittsburgh approach seems more natural.

On the other hand, the Michigan approach might be more natural in other kinds of data mining tasks. The Pittsburgh approach directly takes into account rule interaction when computing the fitness function of an individual. However, this approach leads to syntactically-longer individuals, which tends to make computation expensive. In addition, it may require some modifications to standard genetic operators to cope with relatively complex individuals.

By contrast, in the Michigan approach the individuals are simpler and syntactically shorter. This tends to reduce the time and to simplify the design of genetic operators. However, this advantage comes with a cost. [12]

So far we have seen that an individual of a Genetic Algorithm can represent a single rule or several rules, but we have not said yet how the rule(s) is(are) encoded in the genome of the individual. We now turn to this issue. To follow our discussion, assume that a rule has the form "IF *cond*1 AND ... AND *cond*n THEN *class* = *ci*", where *cond*1 ... *cond*n are attribute-value conditions (e.g. *Sex* = "M") and *ci* is the class predicted by the rule. We divide our discussion into two parts, the representation of the rule antecedent (the "IF" part of the rule) and the representation of the rule consequent (the "THEN" part of the rule).

Hence, this procedure effectively encouraging the evolution of several different rules, each of them covering a different part of the data space.

### 4.1.2 Generalizing/Specializing Crossover

The basic idea of this kind of crossover is to generalize or specialize a given rule, depending on whether it is currently overfitting or underfitting the data, respectively. An Underfitting is the dual situation, in which a rule is covering too many training examples, and so should be specialized. This crossover operator can be implemented as the logical OR and the logical AND, respectively. [13]

### 4.1.3 Generalizing/Specializing-Condition Operator

In the previous subsection we saw how the crossover operator can be modified to generalize/ specialize a rule. However, the generalization/specialization of a rule can also be done in a way independent of crossover. Suppose, e.g., that a given individual represents a rule antecedent with two attribute-value conditions, as follows - again, there is an implicit logical AND connecting the two conditions in (a):

(*Age* > 25) (*Marital_Status* = “single”). (a)

We can generalize, say, the first condition of (a) by using a kind of mutation operator that subtracts a small, randomly-generated value from 25. This might transform the rule antecedent (a) into, say, the following one:

(*Age* > 21) (*Marital_Status* = “single”). (b)

Rule antecedent (b) tends to cover more examples than (a), which is the kind of result that we wish in the case of a generalization operator. Another way to generalize rule antecedent (a) is simply to delete one of its conditions. This is usually called the drop condition operator in the literature. Conversely, we could specialize the first condition of rule antecedent (a) by using a kind of mutation operator that adds a small, randomly-generated value to 25[14].

To simplify our discussion, throughout this subsection we will again assume that the Genetic Algorithm follows the Michigan approach .

# 5. EVALUATING THE QUALITY OF A RULE

Let a rule be of the form: IF A THEN C, where A is the antecedent (a conjunction of conditions) and C is the consequent (predicted class), as discussed earlier. A very simple way to measure the predictive accuracy of a rule is to compute the so-called confidence factor (CF) of the rule, defined as:

$$CF = |A \& C| / |A| \qquad (1)$$

Where |A| is the number of examples satisfying all the conditions in the antecedent A and |A & C| is the number of examples that both satisfy the antecedent A and have the class predicted by the consequent C. For instance, if a rule covers 10 examples (i.e. |A| = 10), out of which 8 have the class predicted by the rule (i.e. |A&C| = 8) then the CF of the rule is CF = 80%. Unfortunately, such a simple predictive accuracy measure favors rules overfitting the data. For instance, if |A| = |A & C| = 1 then the CF of the rule is 100%. However, such a rule is most likely representing an idiosyncrasy of a particular training example, and probably will have a poor predictive accuracy on the test set. A solution for this problem is described next. The predictive performance of a rule can be summarized by a 2 x 2 matrix, sometimes called a confusion matrix

We can now measure the predictive accuracy of a rule by taking into account not only its CF but also a measure of how “complete” the rule is, i.e. the proportion of examples is, having the predicted class C that is actually covered by the rule antecedent. The rule completeness measure is computed by the formula:

$$\text{Completeness} = |A \& C| / |C| \qquad (2)$$

Where |C| is the number of examples satisfying all the conditions in the consequent C and |A & C| is the number of examples that both satisfy the antecedent A and have the class predicted by the consequent C. For instance, if a rule covers 10 examples (i.e. |C| = 10), out of which 5 have the class predicted by the rule (i.e. |A&C| = 5) then the Completeness of the rule is = 50%.The rule interestingness measure is computed by the formula:

$$\text{Interestingness} = |A\&C| - (|A|*|C|)/N \qquad (3)$$

Where |A&C| is the number of examples that both satisfy the antecedent A and have the class predicted by the consequent C minus the product of |A| is the number of examples satisfying all the conditions in the antecedent A and |C| is the number of examples satisfying all the conditions in the consequent. The Following expression can be used to quantify the comprehensibility of an association rule

$$\text{Comprehensibilty}=\log(1+|C|)/\log(1+|A\&C|) \qquad (4)$$

Where |A&C| is the number of examples that both satisfy the antecedent A and have the class predicted by the consequent C and |C| is the number of examples satisfying all the conditions in the consequent.

# 6. THE PROPOSED ALGORITHM

Our approach works as follows:

1. Start
2. Load a sample of records from the database that fits into the memory.
3. Categorical attributes are transformed into asymmetric binary variables by introducing as many new “items” as the number of distinct attribute-value pairs.
4. Apply Apriori algorithm to find the frequent itemsets with the minimum support. Suppose S is set of the frequent item set generated by Apriori algorithm.
5. Apply Genetic Algorithm for generation of all rules.
6. Set Q=Ø where Q is the output set, which contains the entire association rule.
7. Set the Input termination condition of genetic algorithm.
8. Represent each frequent itemset of S as binary encoding, string using the combination of representation specified in method above.
9. Select the two members (string) from the frequent item set using Roulette Wheel sampling method.
10. Apply Genetic Algorithm operators on the selected members (string) to generate the association rules.
11. Find confidence factor, completeness, comprehensibility and interestingness for x=>y each rule.
12. If generated rule is better than previous rule then
13. Set Q = Q U {x =>y}
14. If the desired number of generations is not completed, then go to Step 3.
15. Decode the chromosomes in the final stored generations and get the generated rules.
16. Select rules based on comprehensibility and interestingness
17. Stop.

# 7.RESULTS AND DISCUSSION

Experiments were conducted using real-world Zoo dataset. For brevity, the data used is of a categorical nature. The Zoo database contains 101 instances corresponding to animals and 18 attributes. The attribute corresponding to the name of the animal was not considered in the valuation of the a1gorithm. This was mainly due to its descriptive nature. The attribute from the datasets that were used for analysis include: hair [H], eggs [E], milk [M], airborne [A], predator [P], backbone[B], breathes [BR], venomous [V], tail [T], domestic [D], fins[FN] , aquatic[AQ] and feathers [F].Default values of the parameters are: Population size = 50, Mutation rate = 0.5, Crossover rate =0.8.The stopping criterion used is the non evolution of the archive during 10 generations, once the minimal number of generations has been over passed.

In the following table are the results of the experiments conducted. In these tests, different predictions were made by combining different attributes to determine a result.

Table. 4 The following are results from the given data

| Rule no. | Discovered rule (antecedent->consequent) | Predictive Accuracy/Confidence Factor | Completeness | Interesting-ness | Compreh-ensibility |
|---|---|---|---|---|---|
| 1 | BΛPΛV→˥D | 25 | 1.1905 | 0.32 | 0.9566 |
| 2 | FΛEΛA→P | 6.429 | 1.667 | 1.78 | 0.8703 |
| 3 | MΛFNΛT→BR | 25 | 1.19 | 0.32 | 0.957 |
| 4 | AQΛF→˥H | 16.67 | 2 | 1.5 | 0.829 |
| 5 | FΛE→AQ | 4.3 | 2.688 | 0.2 | 0.749 |

As it is indicated in the results table, overall the discovered rules are quite interesting. we can find out the measurement of association rules using genetic algorithm. Here, we have got some association rules which optimal according their interestingness and comprehensibility. If we study the result then, we can select 12 attributes among 18 attributes.

## 8.CONCLUSION & FUTURE WORK

We have deal with an association rule mining problem for optimized association rules. The frequent itemsets are generated using the Apriori association rule mining algorithm. After all rule generation, GA are apply to optimize generate rule. The use of a multi-objective evolutionary framework for association rule mining offers a tremendous flexibility to exploit in further work. In this present work, we have used a Pareto based genetic algorithm to solve the multi-objective rule mining problem using measures— comprehensibility and the predictive accuracy. We adopted a variant of the Michigan approach to represent the rules as chromosomes, where each chromosome represents a separate rule. This approach can be worked with numerical valued attributes as well as categorical attributes. This work is able to select all useful attributes for any sort of dataset. The results reported in this paper are very promising since the discovered rules are of optimized rules. We can use other technique to minimize the complexity of the genetic algorithm. No threshold value is used here. We can also use other evolutionary algorithm for optimization of association rules. Moreover, we tested the approach only with the numerical and categorical valued attributes. It must be tested with the continuous attributes also.

**Authors**

Malar Vizhi , working as professor in Department of Computer Application in Hindu College, Pattabiram, Chennai. She received her degrees, M.Phil.(C.S.) from Alagappa University and M.C.A. from Madras University, India. Her area of research includes Data Mining, Computer networks.

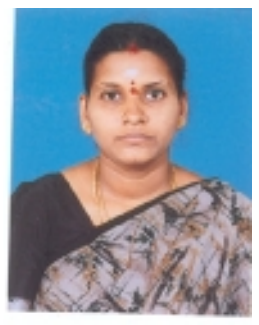

Bhuvaneswari, working as Assistant Professor in Government .Arts and Science College in Department of Computer Science. She received her Ph.D from SCSVMV University, Enathur . Her area of research is Data Mining. She published papers in national and international journals, national and international conferences.

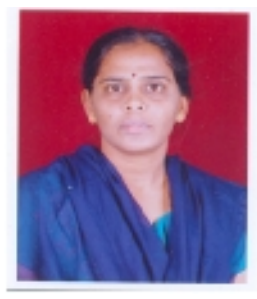